\documentclass[runningheads]{llncs}

\usepackage[T1]{fontenc}

\usepackage[english]{babel}

\usepackage{graphicx}
\usepackage{subfig}
\usepackage{amsmath, amsfonts}

\usepackage{comment}

\usepackage{bm}
\newcommand{\bdt}{\bm{\delta t}}
\newcommand{\bth}{\bm{\vartheta}}
\newcommand{\bL}{\bm{\Lambda}}
\newcommand{\tL}{\tilde{\bm{\Lambda}}}

\usepackage{tikz}
\usetikzlibrary{shapes.geometric,arrows.meta, positioning}

\usepackage{float}

\begin{document}
\title{Addressing prior dependence in hierarchical Bayesian modeling for PTA data analysis I: Methodology and implementation}
\titlerunning{Hierarchical Bayesian Reparametrization for PTA data analysis - I}
%
\author{L. D'Amico\inst{1}\orcidID{0009-0007-4547-9456} \and
E. Villa\inst{2}\orcidID{0000-0003-2203-0254} \and
F. Modica Bittordo\inst{1}\orcidID{0009-0003-1403-4033} \and
A. Barca\inst{1}\orcidID{0009-0007-6350-3798} \and
F. Alì\inst{1}\orcidID{0009-0002-2894-3652} \and
M. Meneghetti\inst{3}\orcidID{0000-0003-1225-7084} \and
L. Naso\inst{1}\orcidID{0009-0002-6495-3321}}
\authorrunning{L. D'Amico et al.}
%
\institute{Koexai Srl, Via Josemaria Escrivà 6, 95125 Catania \and INAF – Istituto di Astrofisica Spaziale e Fisica cosmica di Milano (IASF-MI), Via Alfonso Corti 12, 20133 Milano \and INAF – Osservatorio di Astrofisica e Scienza dello Spazio di Bologna, Via Piero Gobetti 93/3, 40129 Bologna}

\maketitle              
\begin{abstract}
Complex inference tasks, such as those encountered in Pulsar Timing Array (PTA) data analysis, rely on Bayesian frameworks. The high-dimensional parameter space and the strong interdependencies among astrophysical, pulsar noise, and nuisance parameters introduce significant challenges for efficient learning and robust inference. These challenges are emblematic of broader issues in decision science, where model over-parameterization and prior sensitivity can compromise both computational tractability and the reliability of the results.

We address these issues in the framework of hierarchical Bayesian modeling by introducing a reparameterization strategy. Our approach employs Normalizing Flows (NFs) to decorrelate the parameters governing hierarchical priors from those of astrophysical interest. The use of NF-based mappings provides both the flexibility to realize the reparametrization and the tractability to preserve proper probability densities. We further adopt \texttt{i-nessai}, a flow-guided nested sampler, to accelerate exploration of complex posteriors. This unified use of NFs improves statistical robustness and computational efficiency, providing a principled methodology for addressing hierarchical Bayesian inference in PTA analysis.

\end{abstract}
%
%
%
\section{Introduction}

The analysis of Pulsar Timing Array (PTA) data plays a central role in the effort to detect and characterize the Stochastic Gravitational Wave Background (SGWB) at nanohertz frequencies. Evidence of a SGWB has recently been reported by multiple international PTA collaborations \cite{Agazie2023evidence}. PTA sensitivity depends critically on modeling both the SGWB and complex noise processes intrinsic to pulsars and the measurement system. Millisecond pulsars are extremely precise, stable rotators emitting radiation like cosmic lighthouses. A typical PTA model includes physical parameters describing the SGWB spectrum—modeled as a power law with amplitude and spectral index—alongside noise parameters accounting for pulsar-specific contributions, such as white and red noise amplitudes and spectral indices, clock and ephemeris errors, and timing-model parameters \cite{taylor}.  The effects of the SGWB perturbations are encoded in the differencies between the observed Time Of Arrival (TOA) with respect to the theoretical predictions. The  time residuals are given by
\begin{equation} \label{model}
\boldsymbol{\delta t}=M\boldsymbol{\epsilon}+F\boldsymbol{a}+\boldsymbol{n},
\end{equation}
where $\boldsymbol{\epsilon}$ are physical parameters, $\boldsymbol{a}$ Fourier coefficients with design matrix $F$, $M$ the matrix of residual derivatives, and $\boldsymbol{n}$ white noise. The term $F\boldsymbol{a}$ includes correlated and uncorrelated low-frequency processes such as red noise, SGWB, intrinsic spin-noise and dispersion measure. 

Hierarchical Bayesian modeling provides a comprehensive framework for PTA data analysis by allowing the inclusion of priors on noise parameters and subsequent marginalization to estimate posteriors of physical parameters \cite{rutger-hier}, \cite{ensemble}, \cite{goncharov2025readingsignaturessupermassivebinary}. However, posterior inferences are sensitive to prior choices—a well-known problem in Bayesian analysis. Recent PTA studies have explored strategies to mitigate this sensitivity, including parametric uniform priors \cite{rutger-av}, Gaussian priors \cite{ensemble}, and Jeffreys priors for red noise processes \cite{gibbs}.

In this work, we address prior sensitivity in hierarchical PTA modeling through a reparameterization strategy based on parameter orthogonalization \cite{cox-reid-1987}, \cite{tibshirani-wasserman-1994}, \cite{christensen-etal-2006}. The orthogonalization technique proposed in the context of Effective Field Theory in cosmology in \cite{Paradiso_2025} makes use of Generalized Additive Models (GAMs) to decorrelate cosmological and nuisance parameters. As a result the posterior of cosmological parameters is less sensitive to the nuisance prior. We extend this approach by introducing a hierarchical layer for the noise model, placing hyperpriors on noise parameter distributions, and employing Normalizing Flows (NFs) \cite{papaspiliopoulos-etal-2007}, \cite{papamakarios2021flows}, \cite{rezende2015variational}, \cite{Kobyzev_2021} to decorrelate hyperparameters from physical parameters. Inference is performed using \texttt{i-nessai} \cite{nessai}, \cite{Williams_2021}, \cite{Williams_2023}, a flow-guided nested sampling algorithm that efficiently explores high-dimensional, highly correlated parameter spaces, accelerating full Bayesian inference compared to standard Parallel Tempered Markov Chain Monte Carlo (PTMCMC) approaches \cite{nessai-spoke3}.  

This paper is organized as follows. Section~\ref{sec:deco} presents our parameter decorrelation methodology and its implementation via NFs, with Subsection~\ref{subsec:training} focusing on training. Section~\ref{sec:pta} discusses the application to PTA data, including Subsection~\ref{subsec:hierarchical} on the hierarchical Bayesian implementation and Subsection~\ref{sec:testA} on our validation test. Section~\ref{sec:discussion} discusses results, and Section~\ref{sec:concl} summarizes our conclusions.

\section{Parameter decorrelation methodology}\label{sec:deco}

We present here in full detail the construction of the orthogonal reparametrization in a general hierarchical Bayesian setting\footnote{We follow the notation of \cite{ThraneTalbot2019} and \cite{ensemble} for the hierarchical posterior.}. Our framework will be specialized to the PTA context in Section~\ref{sec:pta}. We consider some physical parameters $\bth$ whose prior distribution $\pi(\bth|\bL)$ is parametrized by the hyperparameters $\bL$, which in turn are distributed according to their hyperprior $\pi'(\bL)$, in general different from the distribution $\pi$.
The two-level parameters joint posterior is given by
\begin{equation} \label{post}
\mathcal{P}(\bth,\bL|\bdt) = \frac{\mathcal{L}(\bdt|\bth)\pi(\bth|\bL)\pi'(\bL)}{\mathcal{Z}}\,.
\end{equation}
In the above equation, $\mathcal{L}(\bdt|\bth)$ is the likelihood and depends on the physical parameters only, and $\mathcal{Z}$ is the Bayesian evidence. The hierarchical structure is fully encoded in the parametrized prior term $\pi(\bth|\bL)$, giving the distribution of $\bth$ depending on the hyperparameters $\bL$, which are in turn distributed according to $\pi'(\tL)$. 

The decorrelation procedure is based on projecting out the component of the hyperparameter vector $\bL$ that lies in the subspace spanned by the physical parameters $\bth$. The orthogonal complement to this projection yields the transformed hyperparameter vector $\tL$, which is, by construction, orthogonal to the physical parameters. The reparametrization is thus expressed by the following transformation
\begin{equation}\label{projection}
   {\tL}={\bL}-P_{\bth}\bL=(I-P_{\bth})\bL\,, 
\end{equation}
where
\begin{equation}
    P_{\bth}\equiv\bth(\bth^T\bth)^{-1}\bth^T
\end{equation}
is the projector onto the subspace spanned by $\bth$. Geometrically, this means removing from $\bL$ the component lying along the direction of $\bth$. As a result, the transformed hyperparameters $\tL$ satisfy by construction the orthogonality condition $\bth^T \tL=0$. Our goal is to obtain an equivalent representation, where physical parameters depend on decorrelated hyperparameters $\tL$, which are orthogonal to the physical parameter directions, together with the corresponding distribution of these transformed hyperparameters.
In formulas, we want to obtain the transformation 
\begin{equation}\label{transformation}
    \pi(\bth|\bL)\pi'(\bL) \quad\longrightarrow\quad\pi(\bth|\tilde\bL)\tilde\pi'(\tilde\bL)\,.
\end{equation}
However, the orthogonal projection in Equation~\ref{projection} presents a fundamental problem: it does not admit an inverse. Consequently, the straightforward variable change in Equation~\ref{transformation} is ill-suited. Nevertheless, this difficulty can be circumvented by directly modeling both the distributions $\pi(\tilde\bL)$ and $\pi(\bth|\tilde\bL)$ with NFs~\cite{papamakarios2021flows}, \cite{rezende2015variational}, \cite{Kobyzev_2021}. NFs are a class of generative models that transform complex and potentially singular probability distributions into simpler and tractable ones through a sequence of invertible mappings. The ``normalizing'' attribute refers precisely to their ability to regularize problematic distributions by mapping them to standard distributions, enabling both efficient sampling and exact computations through a collection of subsequent invertible transformations. 

To explain our procedure, we start by sampling from the prior distribution $\pi(\bL)$ and backward through the parameter hierarchy to get draws of $\bth$ and we finally obtain draws of $\tL$ via the projection. That is, we:
\begin{enumerate}
    \item sample $\bL_i\sim\pi(\bL)$ for $i=1, ..., N_{\text{samples}}$;
    \item sample $\bth_i\sim\pi(\bth|\bL)$ for $i=1,...,N_{\text{samples}}$;
    \item get samples of $\tilde\bL$ by transforming the pairs $(\bth_i, \bL_i)$ through the projection $\tilde\bL_i= (I-P_{\bth_i}) \bL_i$ for $i=1...N_{\text{samples}}$
    \item check the orthogonality condition $\bth_i^T \tL_i=0$ for the samples.
\end{enumerate}
Once we have $N_{\text{samples}}$ of the triple $(\bth_i, \bL_i, \tL_i)$, we employ two complementary NFs: the first, that we call \textit{Push-forward}, learns from the draws $(\tilde\bL_i)$ the distribution $\pi'(\tilde{\bm{\Lambda}})$ of the decorrelated hyperparameters and then the second, that we call \textit{Pull-backward}, learns the conditional distribution $\pi(\bm{\vartheta}|\tilde{\bm{\Lambda}})$. This yields the required quantities for the transformation in Equation~\ref{transformation}: by taking advantage of NFs we approximate the projection in Equation \ref{projection} and regularize its inverse, while remaining compatible with the orthogonalization and the hierarchical structure of priors and hyperpriors. In the following, we describe in more detail the two algorithms introduced above.

\textbf{Push-forward Normalizing Flow (PF-NF)}: the first component models the distribution $\pi'(\tilde{\bm{\Lambda}})$ of the orthogonalized hyperparameters.  
It is implemented as a Masked Autoregressive Flow (MAF) \cite{papamakarios2017masked}, with three transformation blocks, each consisting of masked affine autoregressive transforms with 32 hidden units. To avoid artifacts from a fixed variable ordering, random permutations are introduced between successive blocks. The base distribution can be flexibly chosen as either a standard Gaussian, $\mathcal{N}(0, \mathbf{I})$, or a uniform distribution, $\mathcal{U}([0,1])$.  
The MAF architecture ensures exact invertibility with a tractable Jacobian computation in $\mathcal{O}(m)$ time, a feature that is essential for both efficient sampling and accurate density evaluation.

\textbf{Pull-backward Conditional Normalizing Flow (PB-CNF)}: the second component is a conditional NF \cite{trippe2018conditional}, that learns the distribution of physical parameters given the hyperparameters, i.e.\ $\pi(\bm{\vartheta}|\tilde{\bm{\Lambda}})$.  
This is realized as a conditional MAF where the context, namely the decorrelated hyperparameters $\tilde{\bm{\Lambda}}$, is injected at each transformation layer. The architecture is composed of three conditional masked affine transforms with shared hyperparameters across layers.  
This conditional structure allows the flow to capture the hierarchical dependency between physical parameters and hyperparameters, while preserving by construction the orthogonality constraint between the two spaces.

\begin{figure}[htbp]
    \centering
    \includegraphics[width=0.75\linewidth]{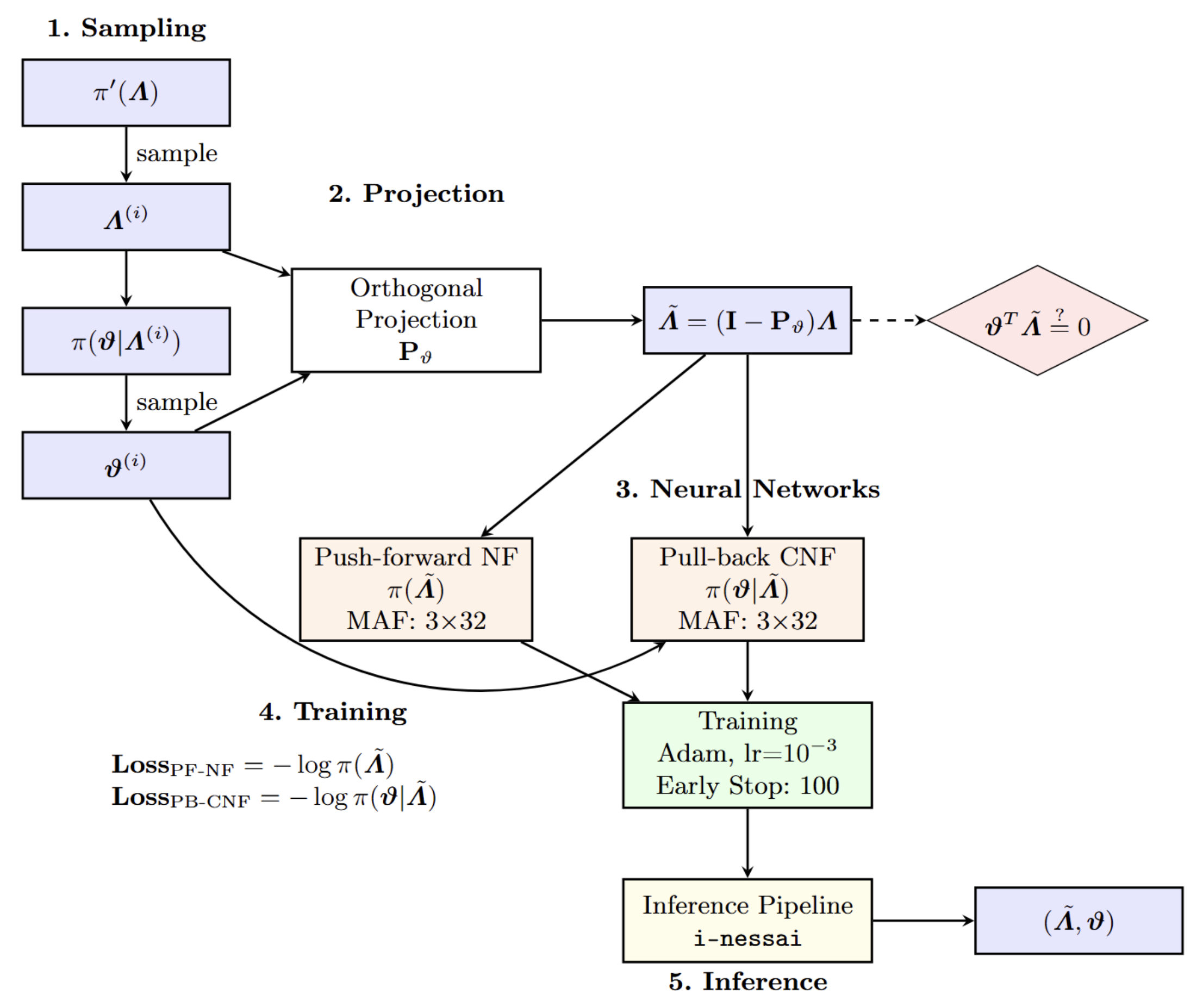}
    \caption{Pipeline for hierarchical decorrelation: sample ($\bm{\theta},{\bm\Lambda}$), project to $\tilde{\bm{\Lambda}}$, learn $\pi\left(\bm{\Lambda}\right)$ and $\pi\left(\bm{\theta}|\bm{\Lambda}\right)$ with NFs, then infer with i-nessai.\\}
    \label{fig:repara_diagram}
\end{figure}

A diagram of the full procedure is shown in Fig.~\ref{fig:repara_diagram}.

\subsection{Push-forward and Pull-backward NFs training}\label{subsec:training}

We trained the two NFs --- the PF-NF and the PB-CNF --- on a dataset consisting of $N_{\text{samples}}=20000$ realizations generated from the priors of $\tilde\Lambda$ and $\theta$. As a preliminary step, all previous samples were rescaled to the interval $[0, 1]$, a transformation that generally improves the convergence of NFs. Although the available hardware would have allowed full-batch training, we adopted a mini-batch strategy with a batch size 256. This choice introduces stochasticity into the optimization process, helping the training escape poor local minima and improving the overall robustness of convergence.

\begin{figure}[htbp]
  \centering
  \subfloat[NF-PF loss history\label{fig:unconditional}]{
    \includegraphics[width=0.45\linewidth]{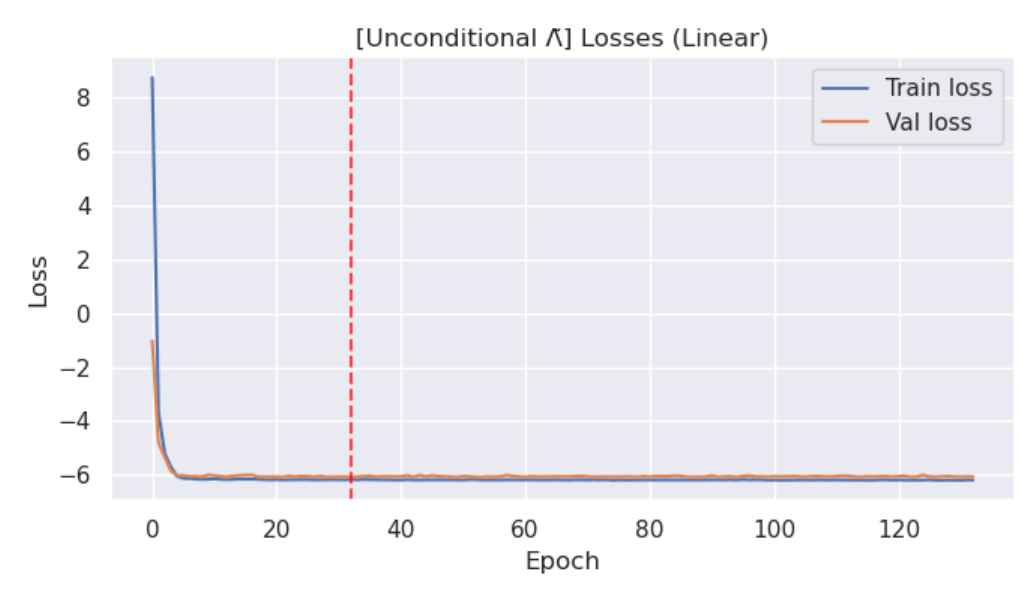}
  }\hfill
  \subfloat[PB-CNF loss history\label{fig:conditional}]{
    \includegraphics[width=0.45\linewidth]{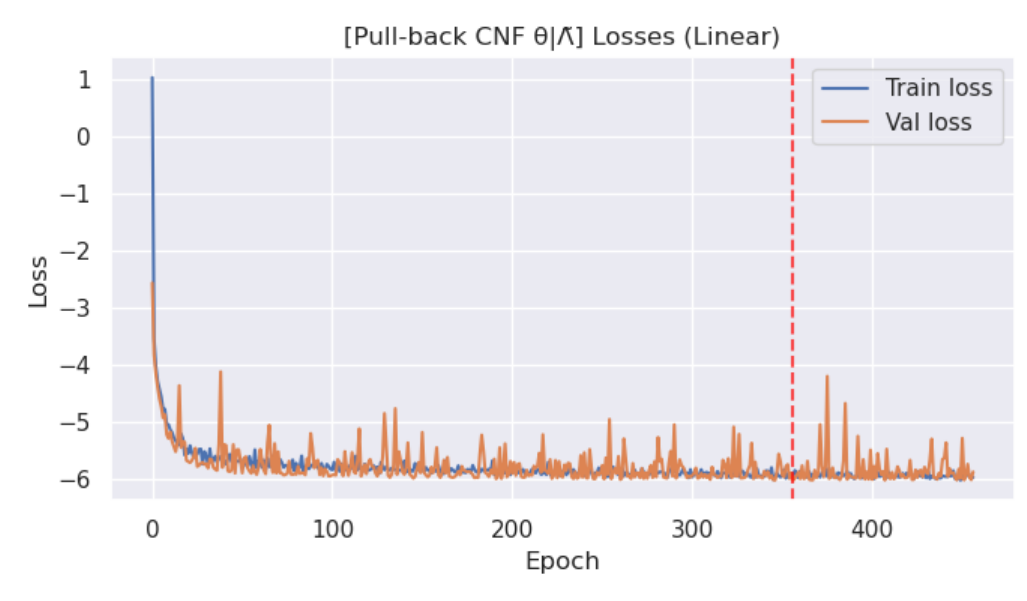}
  }
  \caption{Training (blue) and validation (orange) losses for both unconditional (\ref{fig:unconditional}) and conditional networks (\ref{fig:conditional}) on a linear scale. The vertical red dashed line indicates the selected early stopping epoch. Overall, the losses converge to low and stable values, confirming efficient and robust training.\\}
  \label{fig:training}
\end{figure}

The dataset was further split into training and validation subsets, with 10\% of the samples reserved for validation. This separation serves two purposes: (i) it provides an unbiased evaluation of model performance on unseen data, and (ii) it enables the adoption of an early-stopping criterion, ensuring that the selected model corresponds to the minimum validation loss and reducing the risk of overfitting. The loss function used throughout training is the standard log-likelihood objective common to NFs optimization. In particular, the PF-NF is optimized by maximizing the log-likelihood of the decorrelated hyperparameters:

\begin{equation}
   \textbf{Loss}_{\text{PF-NF}} = -\frac{1}{N_{\text{batch}}} \sum_{i=1}^{N_{\text{batch}}} \log \pi_{\text{NF}}(\tilde{\bm{\Lambda}}^{(i)}) \, ,
\end{equation}
while the PB-CNF maximizes the conditional log-likelihood of the physical parameters given the hyperparameters:
\begin{equation}
    \textbf{Loss}_{\text{PB-CNF}} = -\frac{1}{N_{\text{batch}}} \sum_{i=1}^{N_{\text{batch}}} \log \pi_{\text{CNF}}(\bm{\vartheta}^{(i)}|\tilde{\bm{\Lambda}}^{(i)}) \, .
\end{equation}
Figures \ref{fig:training} display the training histories of PF-NF and PB-CNF. In both cases, the loss converges to low and stable values, demonstrating efficient and robust training. The training time takes approximately 9 minutes.

\begin{figure}
  \centering
  \subfloat{
    \includegraphics[width=0.45\textwidth]{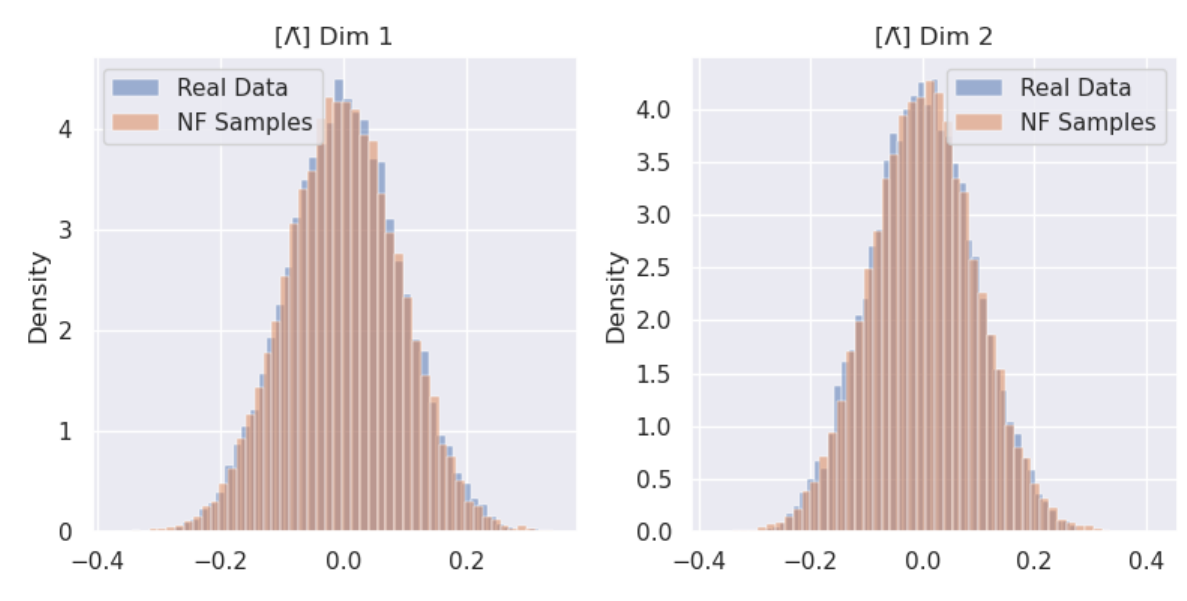}
  }
  \subfloat{
    \includegraphics[width=0.45\textwidth]{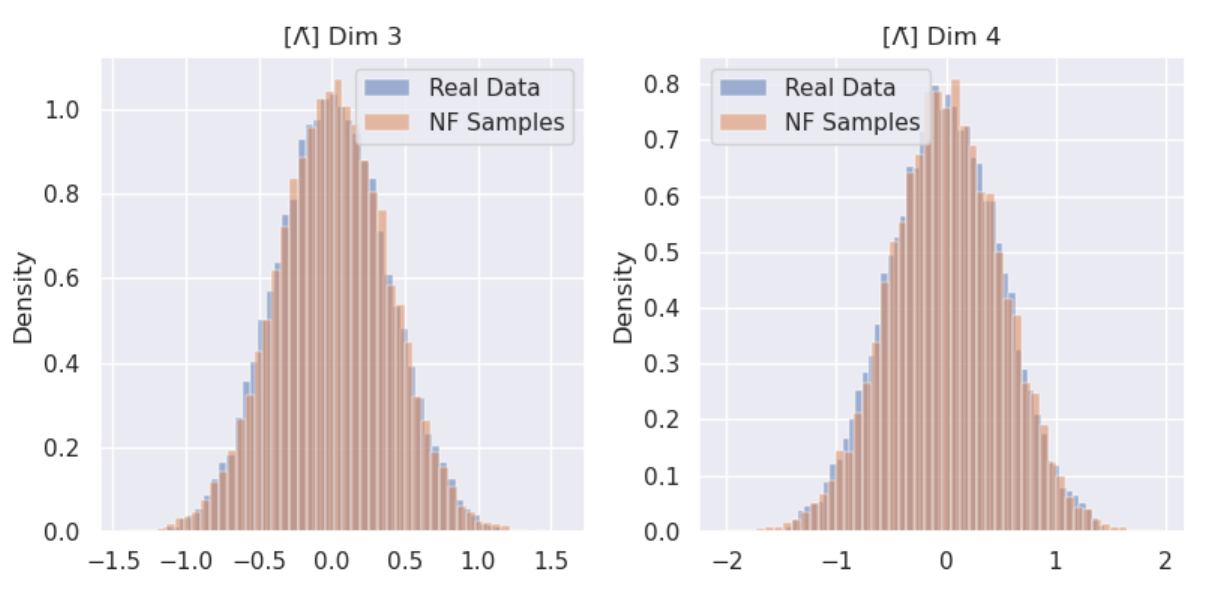}
    }
  \vspace{1mm}
  \subfloat{
    \includegraphics[width=0.45\textwidth]{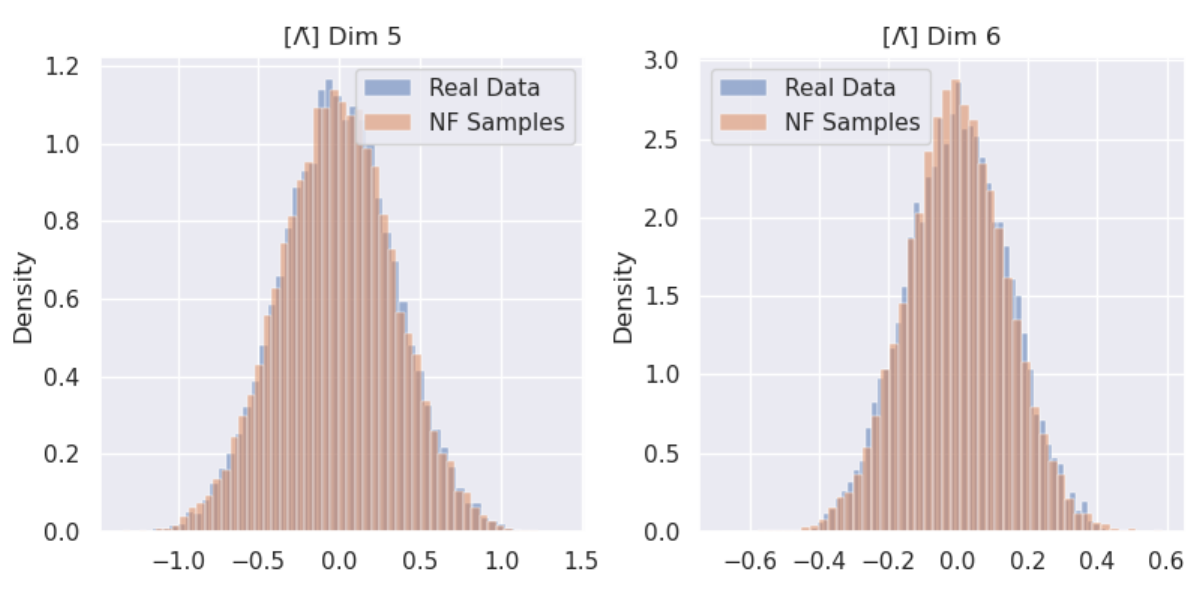}
  }
  \subfloat{
    \includegraphics[width=0.45\textwidth]{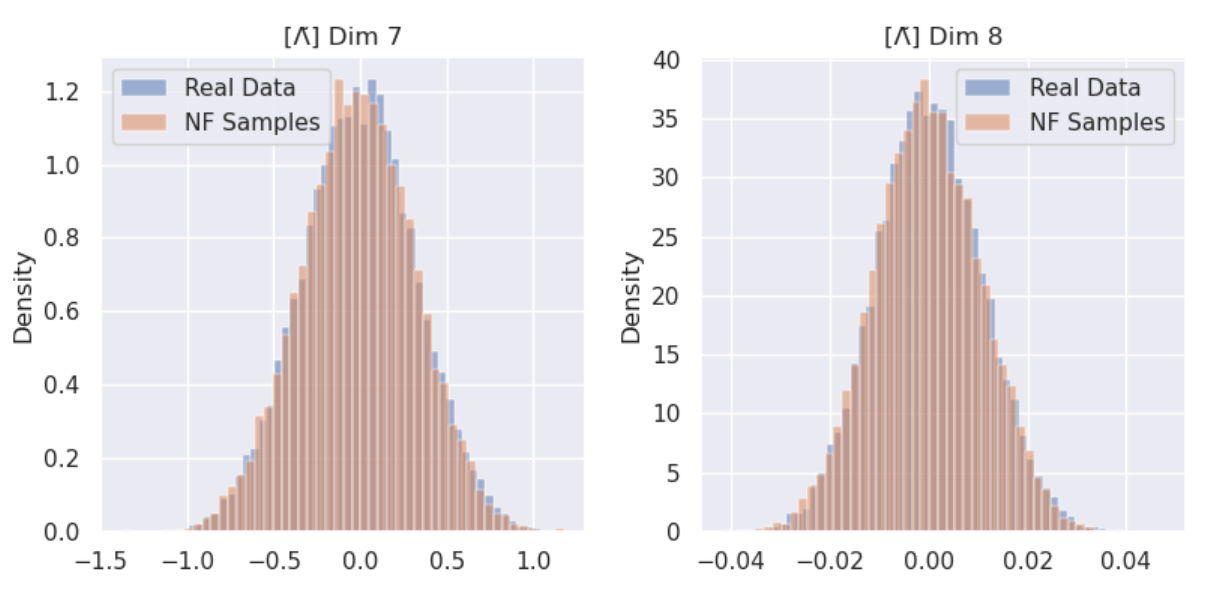}
  }
    \caption{
    Real data PDF (blue) vs PF-NF samples (orange) for eight $\tilde \Lambda$ dimensions; close agreement shows the model reproduces the target distribution.
    }
    \label{fig:distributionPFNF}
\end{figure}

\begin{figure}[htbp]
    \centering
  \subfloat{
    \includegraphics[width=0.45\textwidth]{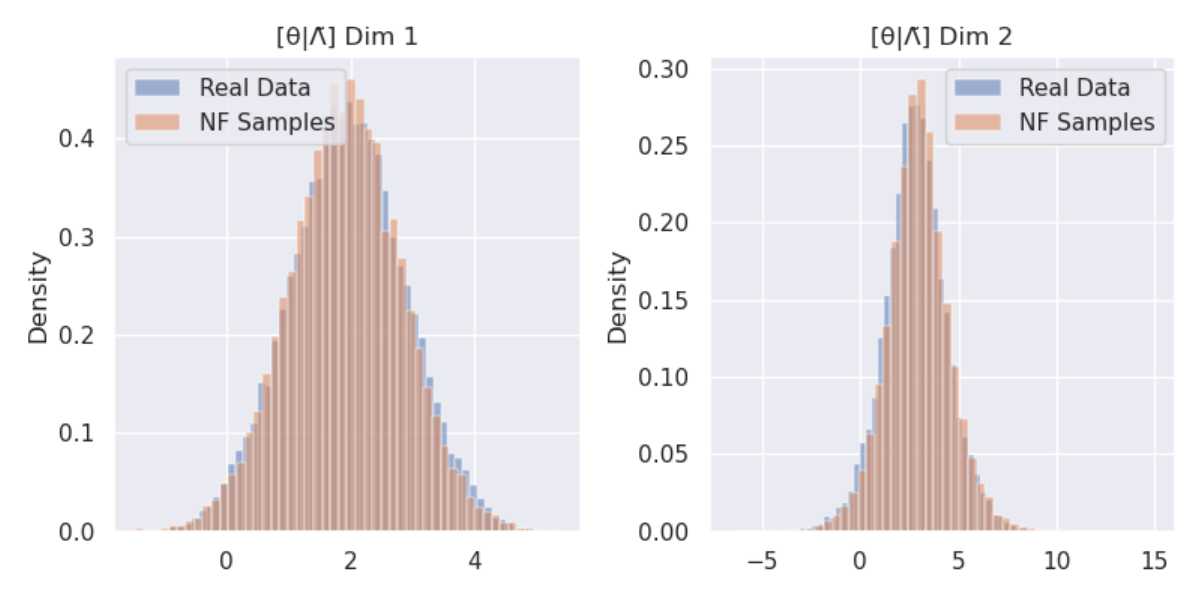}
  }
  \subfloat{
    \includegraphics[width=0.45\textwidth]{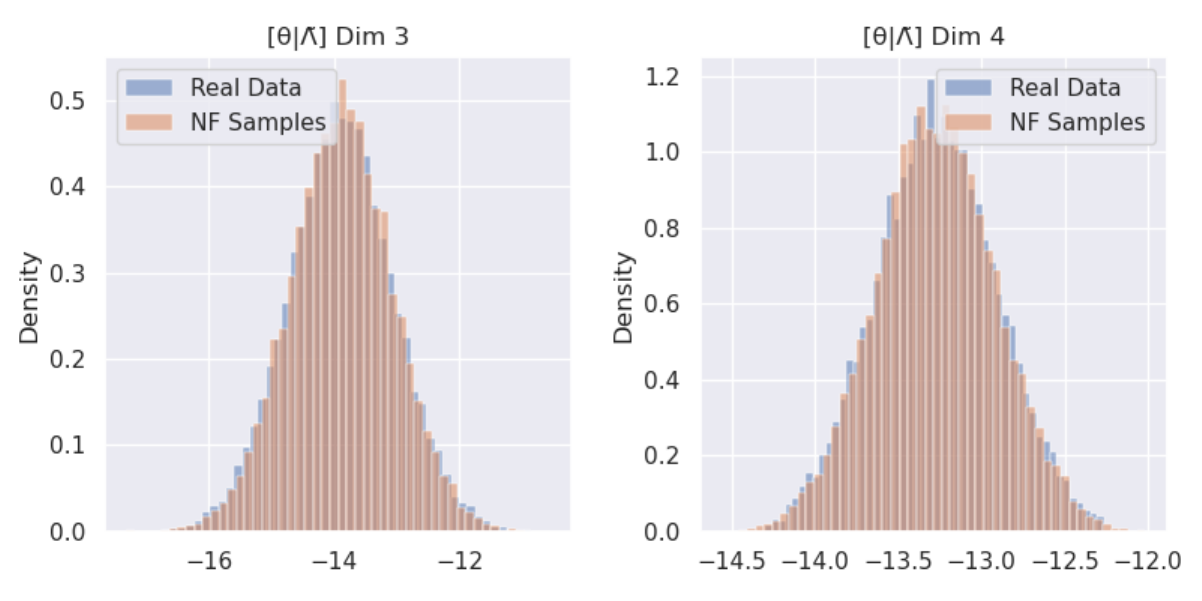}
  }
    \caption{
    Real data PDF (blue) vs PB-CNF samples (orange) for four $\vartheta$ dimensions; close agreement indicates the model reproduces the target distribution.\\
    }
    \label{fig:distributionPLCNF}
\end{figure}

As an additional diagnostic, we compare samples generated from the trained flows with the validation data drawn from the priors. Figures \ref{fig:distributionPFNF} and \ref{fig:distributionPLCNF} show the resulting distributions, where the generated samples visually reproduce the prior distributions with good fidelity. This agreement indicates that both flows have successfully captured the target probability structure.

\section{Application to PTA Data Analysis}\label{sec:pta}

Our PTA inference pipeline relies on two complementary tools. The first is \texttt{Enterprise} (Enhanced Numerical Toolbox Enabling a Robust PulsaR Inference SuitE) \cite{enterprise}, a Python-based software package that has become the de facto standard for PTA data analysis. It offers a modular architecture where pulsar noise and gravitational wave models are defined as components of a probabilistic model. This enables combining timing models, various stochastic noise processes, and common signals across the array, such as the SGWB.
To perform the Bayesian inference we use \texttt{i-nessai} (Nested Sampling with Artificial Intelligence), a nested sampling algorithm that incorporates NFs. During the run, a NF is trained on the live points, allowing the sampler to capture complex posterior geometries and generate new samples following the likelihood contours. This greatly improves efficiency in high-dimensional correlated spaces, reduces the number of likelihood evaluations, and makes it particularly effective for PTA data analysis.
In our reparameterized framework, the geometry of the posterior is simplified but remains non-trivial, and thus benefits directly from the flow-based sampling strategy.

To assess the validity of our reparameterization strategy, we apply our framework to the noise parameter inference of a single pulsar whose timing residuals are simulated from the DR2new release of the European Pulsar Timing Array dataset \cite{EPTA:DR2_I}. According to the standard prescription, we start by considering the PTA likelihood in the form that it assumes after the analytical marginalization over the timing model parameters, which is fully implemented in \texttt{Enterprise}. Moreover, in a good approximation, the white noise components are independent of other noise terms. Therefore, we fix the white noise parameters to their maximum-likelihood values and simply ignore them for the inference. Among the noise processes we account for the intrinsic Red Noise (RN) and the Dispersion Measure (DM) noise. Both are modeled as power laws with two parameters each: the spectral index $\gamma$ and the $\log_{10}$ of the amplitude $A$. Let us finally note that the SGWB signal is absent in a single-pulsar analysis, since it manifests itself as a correlated signal across a collection of pulsars. In the notation of Section~\ref{sec:deco} we thus have in total four components in the vector of physical parameters $\bth$. Parametrized conditional priors $\pi(\bth|\bL)$ with hyperpriors $\pi(\bL)$ are placed on the RN and DM parameters.

\subsection{Implementation of the reparametrized hierachical Bayesian framework in \texttt{Enterprise} and \texttt{i-nessai}}\label{subsec:hierarchical}

The implementation of our reparametrized hierachical Bayesian framework in \texttt{Enterprise} and \texttt{i-nessai} is the core of our work. It allows distinguishing between the parameters $\bth$ and the hyperparameters $\bL$ or $\tL$, ensuring that the likelihood is correctly calculated only for the physical parameters $\bth$, according to the hierarchical prescription for the joint posterior of Equation~\ref{post}.
This structure is first implemented by setting the priors in \texttt{Enterprise} to be very wide uniform distributions, i.e. dummy priors, that do not play an effective role in the sampling process. Their only purpose is to register the physical parameters within \texttt{Enterprise} so that the likelihood can be computed through the standard \texttt{get\_lnlikelihood} method, without building a specific extension of the PTA class. This approach ensures that the Bayesian estimate of the posterior is not biased by auxiliary parameters.
All the specifications needed for priors \(\pi(\bth, \bL)\) and hyperpriors \(\pi(\bL)\), together with their probability distributions --- both before and after the reparametrization --- are fully implemented and managed by \texttt{i-nessai}. 
This choice is motivated by practical considerations, as handling the interface on the \texttt{i-nessai} side proved to be much more efficient and flexible. The main characteristics of the implementation in \texttt{i-nessai} are:

\textbf{Parameter separation for likelihood}: only the $\bth$ physical parameters directly affect likelihood, while all other parameters are considered only for posterior and log-prior calculation purposes. Thus, the function \texttt{log\_likelihood} of \texttt{i-nessai} internally calls \texttt{get\_lnlikelihood} of \texttt{Enterprise} only on the correct subset.

\textbf{Sampling strategy}: regardless of whether standard priors before reparameterisation or neural flows after reparameterisation are used, the sampling strategy is defined within the \texttt{from\_unit\_hypercube} method. This consists of computing the Inverse Cumulative Distribution Function (ICDF) for all parameters, thereby mapping unit-hypercube samples to the corresponding prior distributions. The \texttt{nflows} library \cite{nflows2020} provides both the ICDF and the log-PDF evaluations, then \texttt{i-nessai} samples all parameters and calculates log-prior and log-likelihood separately, ensuring a precise estimate of the overall posterior.

The split implementation ensures that the Bayesian inference pipeline correctly evaluates the posterior while keeping \texttt{Enterprise} focused solely on likelihood evaluation. Thus, the integration of \texttt{Enterprise} with \texttt{i-nessai} via this user-defined class provides a useful framework for hierarchical posterior sampling in PTA analysis, while preserving the distinction between physical parameters and hyperparameters. Our implementation is straightforwardly adaptable to bigger or complex PTA datasets, as it deals mainly with the sampling functionalities in \texttt{i-nessai}, without modifying the internal architecture of \texttt{Enterprise}.

\subsection{Validation test}\label{sec:testA}

\begin{figure}[htbp]
  \centering
  \subfloat[
  Gamma posteriors, uniform hyperprior.
  \label{fig:uni_gamma}
  ]{
    \includegraphics[width=0.45\textwidth]{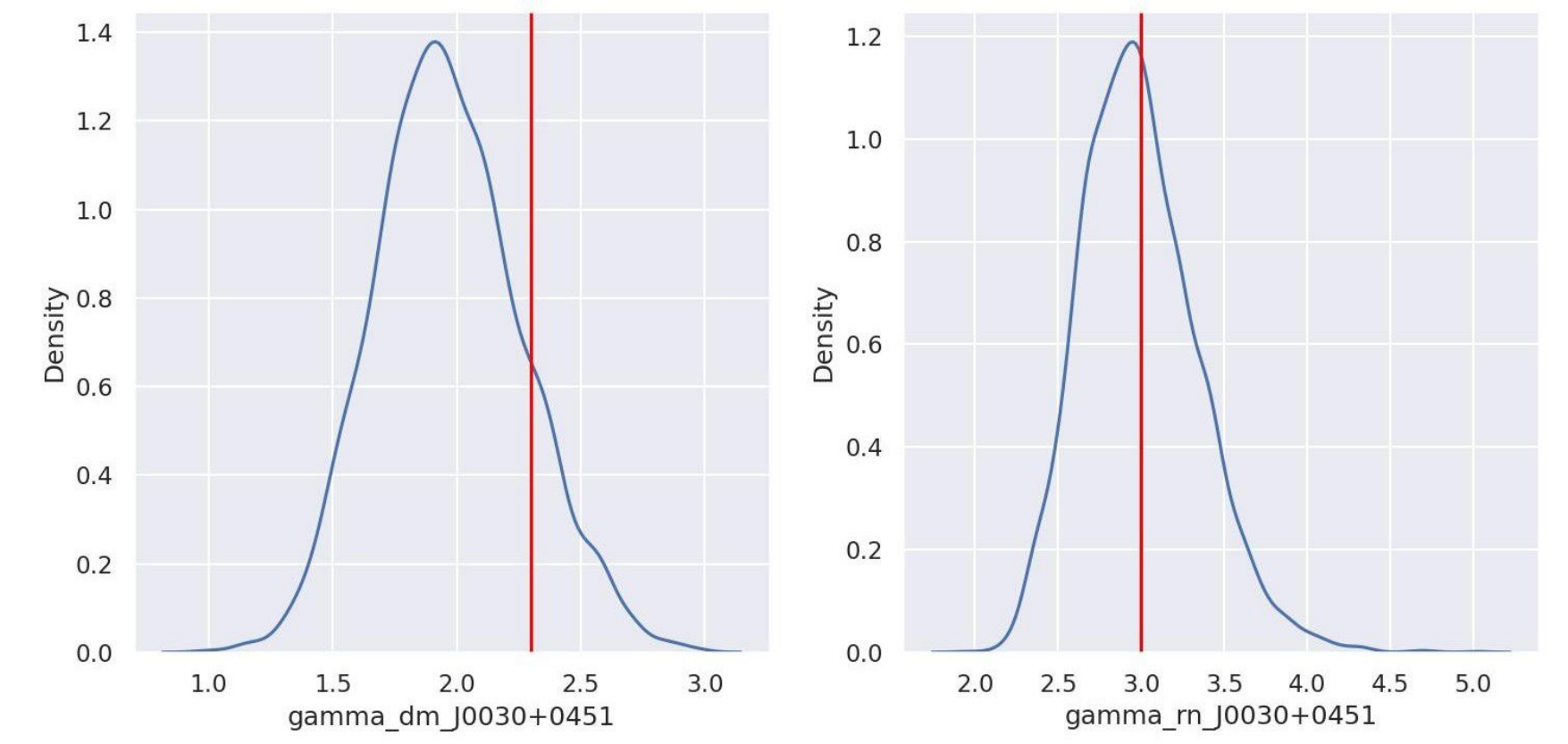}
  }
  \hfill
  \subfloat[
  Amplitude posteriors, uniform hyperprior.
  \label{fig:uni_A}
  ]{
    \includegraphics[width=0.45\textwidth]{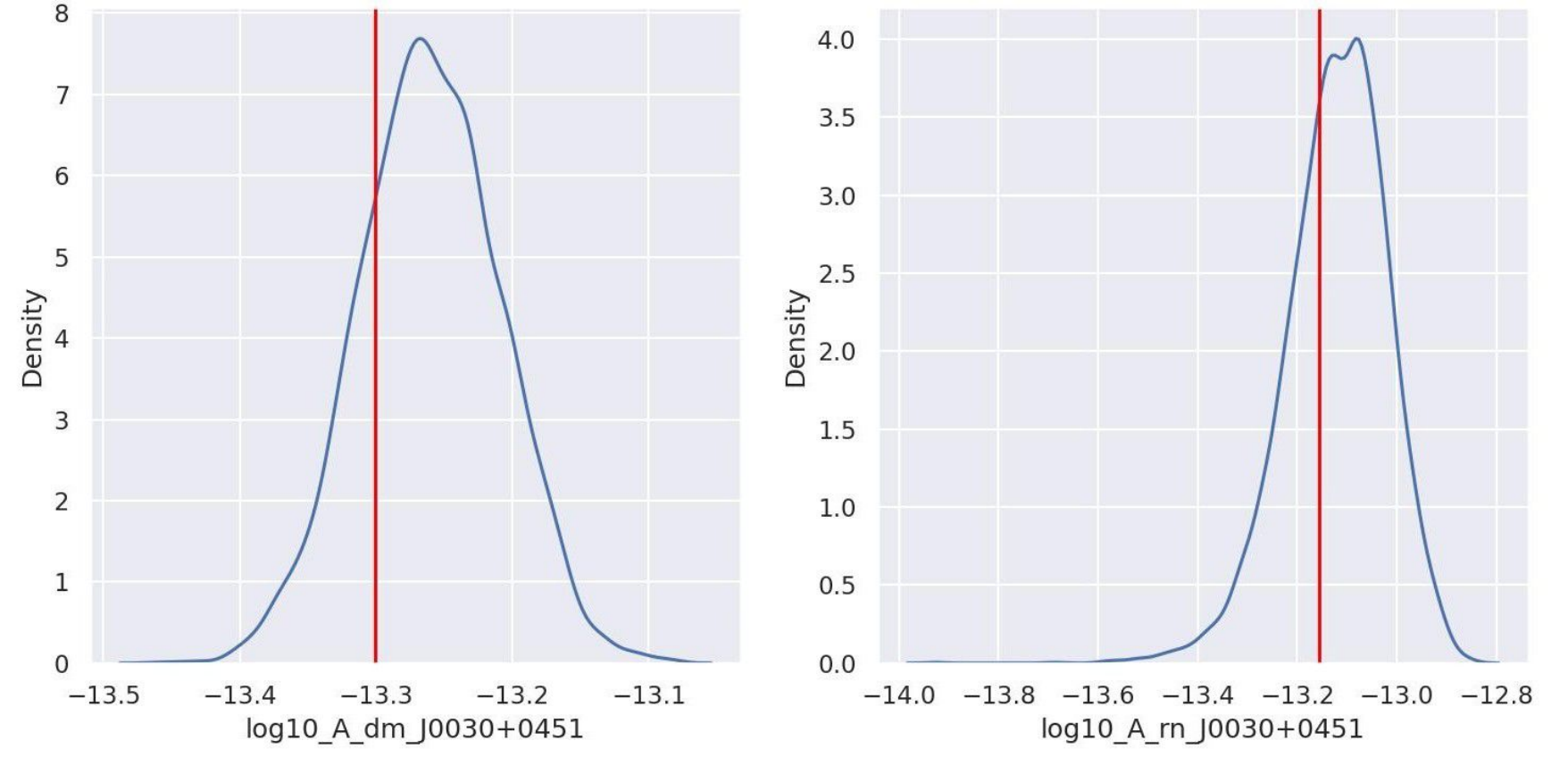}
  }


  \subfloat[
  Gamma posteriors, Gaussian hyperprior.
  \label{fig:gauss_gamma}
  ]{
    \includegraphics[width=0.45\textwidth]{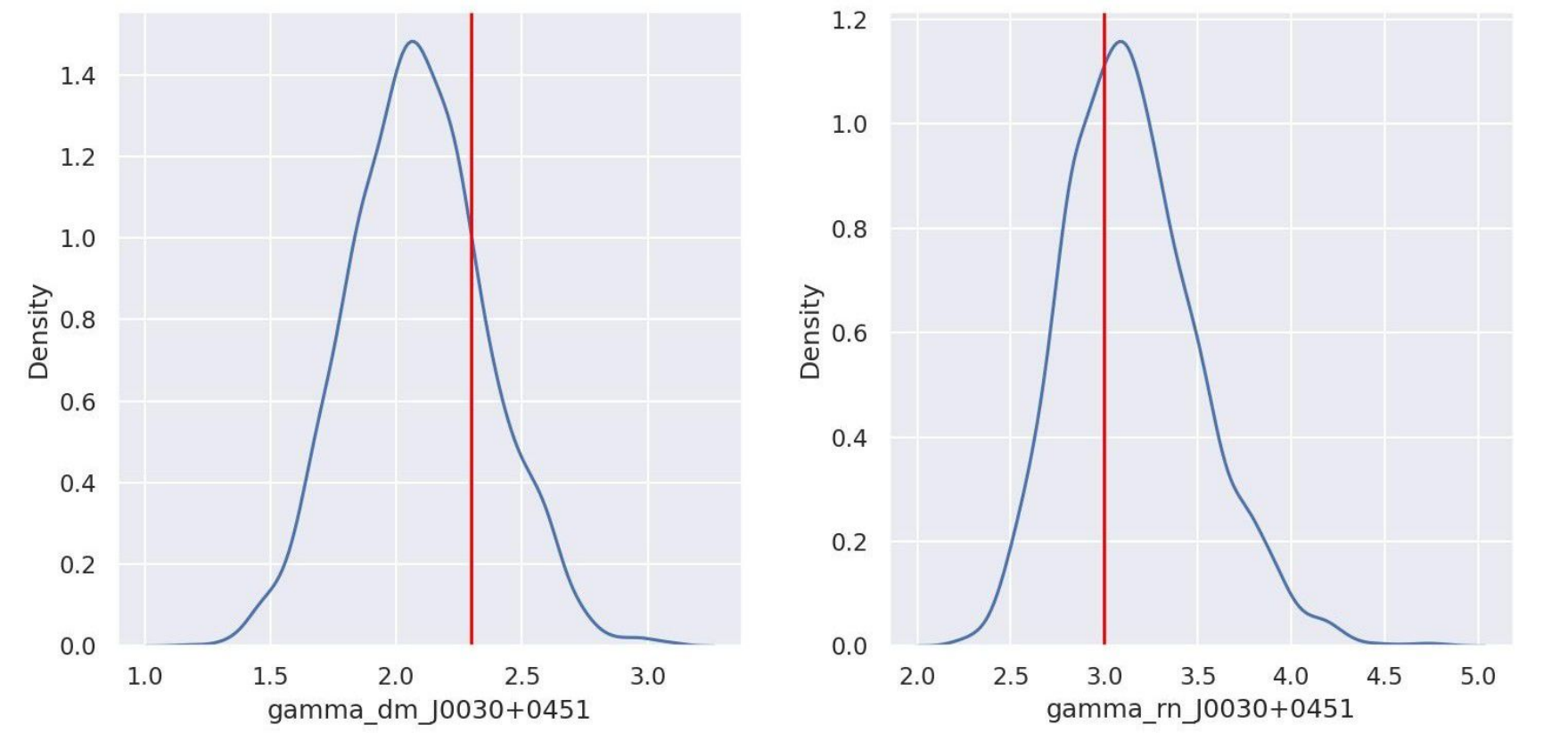}
  }
  \hfill
  \subfloat[%
  Amplitude posteriors, Gaussian hyperprior.
  \label{fig:gauss_A}
  ]{
    \includegraphics[width=0.45\textwidth]{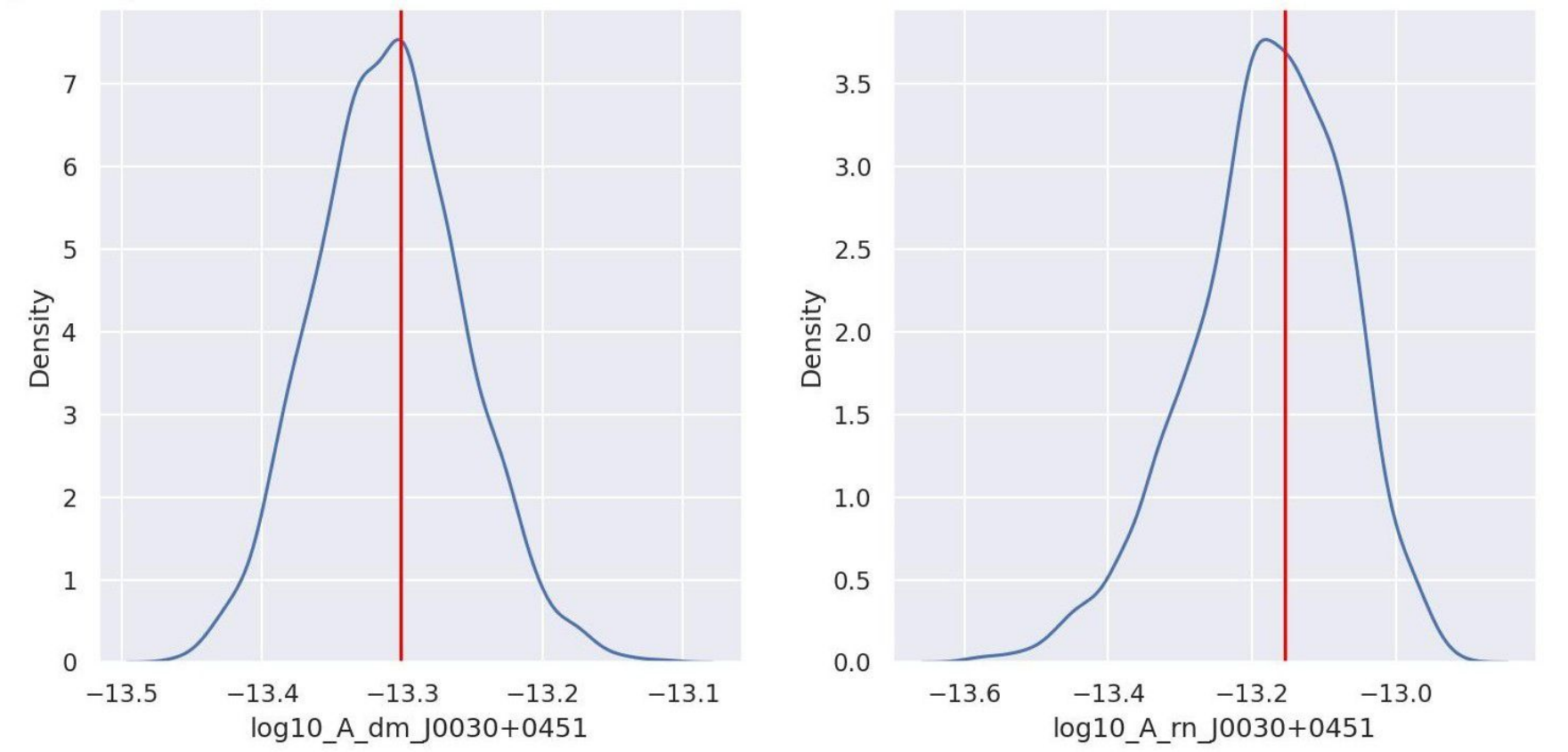}
  }

  \caption{
  Single-pulsar RN and DM posteriors under two hyperpriors: uniform (\ref{fig:uni_gamma}, \ref{fig:uni_A}) vs Gaussian (\ref{fig:gauss_gamma}, \ref{fig:gauss_A}); red lines mark injected values. The hyperprior choice materially affects the inferred parameters (Enterprise likelihood with \texttt{i-nessai}).\\
  }
  \label{fig:comparisonHyperprior}
\end{figure}

In the first place, as a validation test, we consider a single pulsar and we set a Gaussian distribution for the conditional parametrized prior $\pi(\bth|\bL)$. Here the vector $\bth$ has 4 components: $\gamma_{RN}$, $\gamma_{DM}$, $\log_{10}A_{RN}$, and $\log_{10}A_{DM}$. We want to show that the inference of these noise parameters $\bth$ is indeed affected by the choice of the hyperprior on $\bL$, $\pi(\bL)$. In order to quantify this effect we consider two widely-used different hyperprior classes: the Gaussian and the uniform distributions. Each of them adds two additional hyperparameters $\Lambda$: the mean and the standard deviation for the former, and the lower and upper bounds for the latter. As values for the eight hyperparameters in the hyperprior $\pi(\bL)$, we used the results of Table 2 in \cite{ensemble}.
In Figure~\ref{fig:comparisonHyperprior} we report the marginal posterior distributions for the RN and DM noise parameters with uniform (\ref{fig:uni_gamma}, \ref{fig:uni_A}) and Gaussian (\ref{fig:gauss_gamma}, \ref{fig:gauss_A}) hyperprior. The sampling is carried out with \texttt{i-nessai} and we vary the number of live points to optimize posterior accuracy. 
We find that with about 4000 live points, the posterior estimates are sufficiently precise to resolve the differences introduced by different hyperprior choices. Runs with fewer live points reproduce the main pattern, but exhibit significant sampling noise in the tails of the distributions, which is completely consistent with expectations from nested sampling theory. 

To summarize, we remark that the plots in Figure~\ref{fig:comparisonHyperprior} confirm that the detailed shapes of the posteriors depend on the specification of the hyperprior. Furthermore and most importantly, our validation test demonstrates that our implementation of the hierarchical Bayesian modeling of PTA in \texttt{Enterprise}, combined with the use of \texttt{i-nessai} for the sampling, provides a principle-based method for exploring the impact of hyperpriors on the PTA inference.

\section{Discussion}\label{sec:discussion}

To quantitatively assess the effectiveness of our reparametrization procedure, we employ two complementary metrics based on the variance decomposition principle. For any physical parameter $\vartheta_i$ and hyperparameter $\Lambda_j$ (or transformed hyperparameter $\tilde{\Lambda}_j$), the law of total variance states that:
\begin{equation}
\text{Var}(\vartheta_i) = \mathbb{E}[\text{Var}(\vartheta_i|\Lambda_j)] + \text{Var}(\mathbb{E}[\vartheta_i|\Lambda_j])
\end{equation}
where $\mathbb{E}[\text{Var}(\vartheta_i|\Lambda_j)]$ represents the expected conditional variance (the variability in $\vartheta_i$ that remains after accounting for $\Lambda_j$), and $\text{Var}(\mathbb{E}[\vartheta_i|\Lambda_j])$ quantifies the variance in $\vartheta_i$ explained by $\Lambda_j$. Based on this decomposition, we define two key metrics. We define the independence score $\mathcal{I}$ as:
\begin{equation}
(\vartheta_i, \Lambda_j) = \frac{\mathbb{E}[\text{Var}(\vartheta_i|\Lambda_j)]}{\text{Var}(\vartheta_i)}
\end{equation}
which ranges from 0 to 1, with values approaching 1 indicating that $\vartheta_i$ is largely independent of $\Lambda_j$. This metric quantifies the fraction of variance in the physical parameter that is not explained by the hyperparameter, thus measuring the degree of statistical independence.

Conversely, we define the coefficient of determination $R^2$ as:
\begin{equation}
R^2(\vartheta_i, \Lambda_j) = \frac{\text{Var}(\mathbb{E}[\vartheta_i|\Lambda_j])}{\text{Var}(\vartheta_i)} = 1 - \mathcal{I}(\vartheta_i, \Lambda_j)
\end{equation}
which represents the proportion of variance in $\vartheta_i$ that is predictable from $\Lambda_j$. Lower values of $R^2$ indicate better decorrelation. We computed two metrics above for each pair of parameters $(\vartheta_i, \Lambda_j)$ and $(\vartheta_i, \tilde\Lambda_j)$, using kernel ridge regression in the estimation of $\mathbb{E}[\vartheta|\Lambda]$ to capture non-linear dependencies. 

\begin{figure}[htbp]
  \centering
  \subfloat[
  Independence scores (left) and $R^2$ values (right), before projection.
  ]{
    \includegraphics[width=0.45\textwidth]{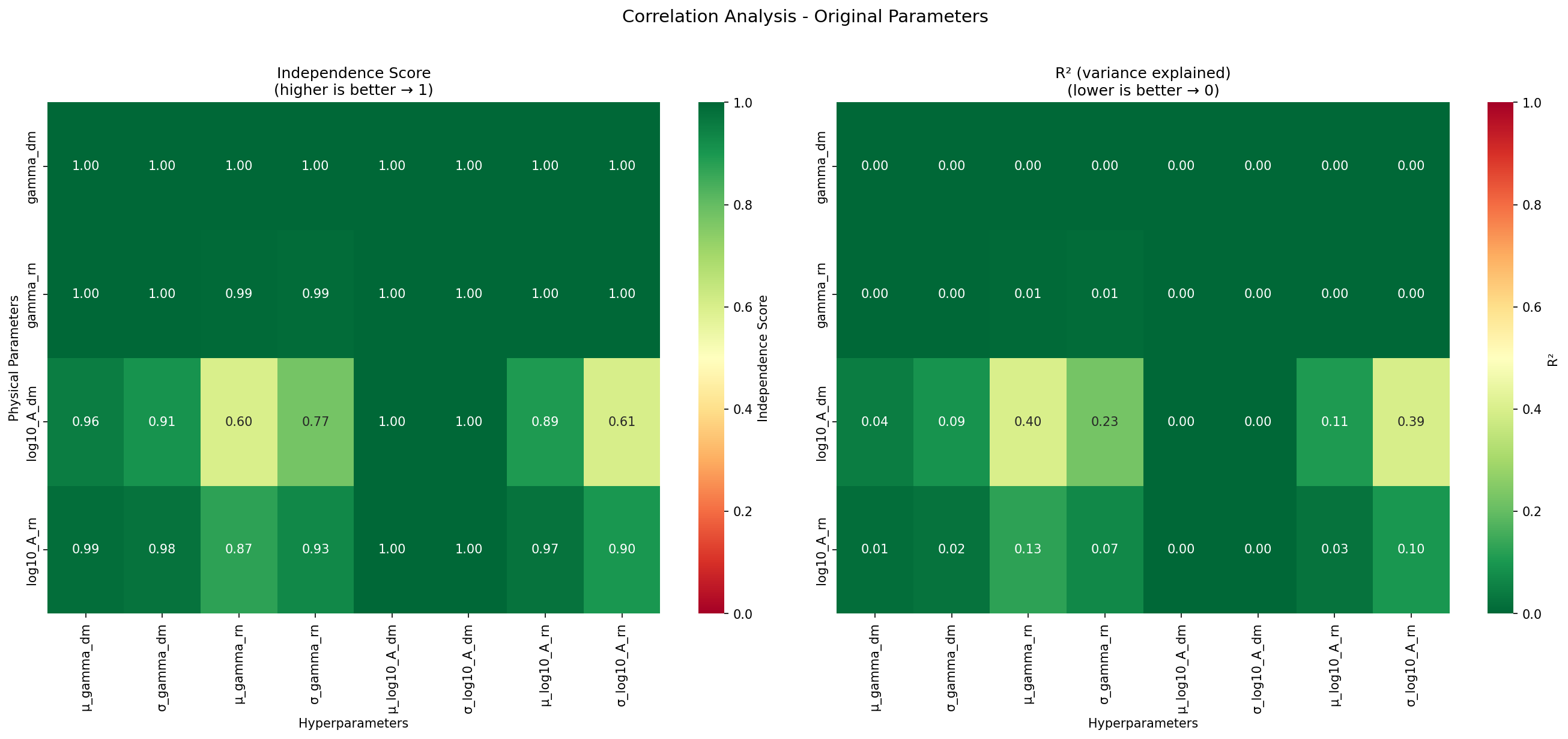} \label{fig:correlation_original}
  }
  \hfill
  \subfloat[
  Independence scores (left) and $R^2$ values (right), after projection.
  ]{
    \includegraphics[width=0.45\textwidth]{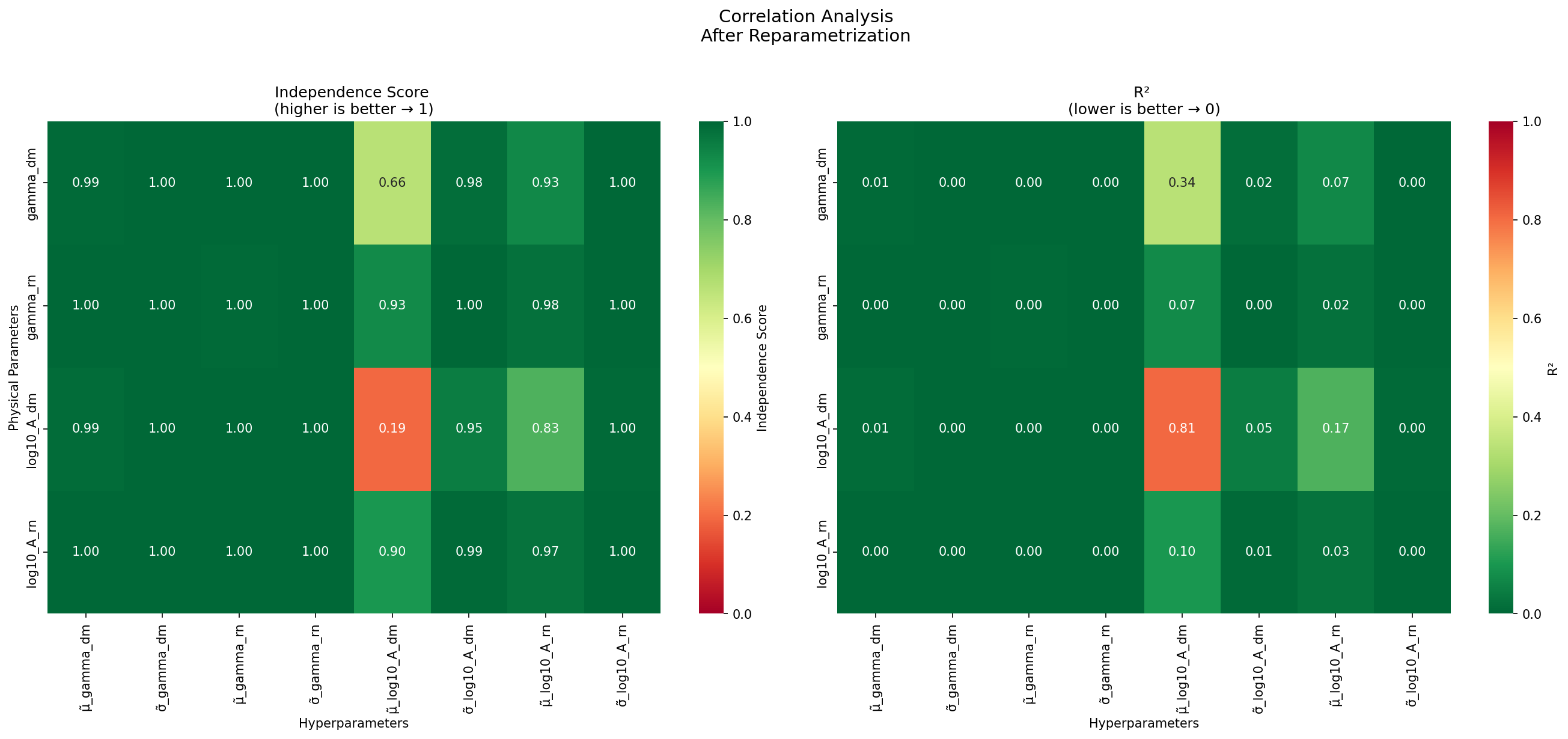}\label{fig:correlation_corrected}
  }
    \caption{
    Independence scores and $R^2$ before (\ref{fig:correlation_original}) and after (\ref{fig:correlation_corrected}) projection: decorrelation increases independence scores and reduces $R^2$. \\
    }  \label{fig:correlations}
\end{figure}

\begin{figure}[htbp]
  \centering
  \subfloat[
  Joint 2D posteriors for $\log_{10}A$ and $\gamma$ (RN, DM) before projection.
  ]{
    \includegraphics[width=0.45\textwidth]{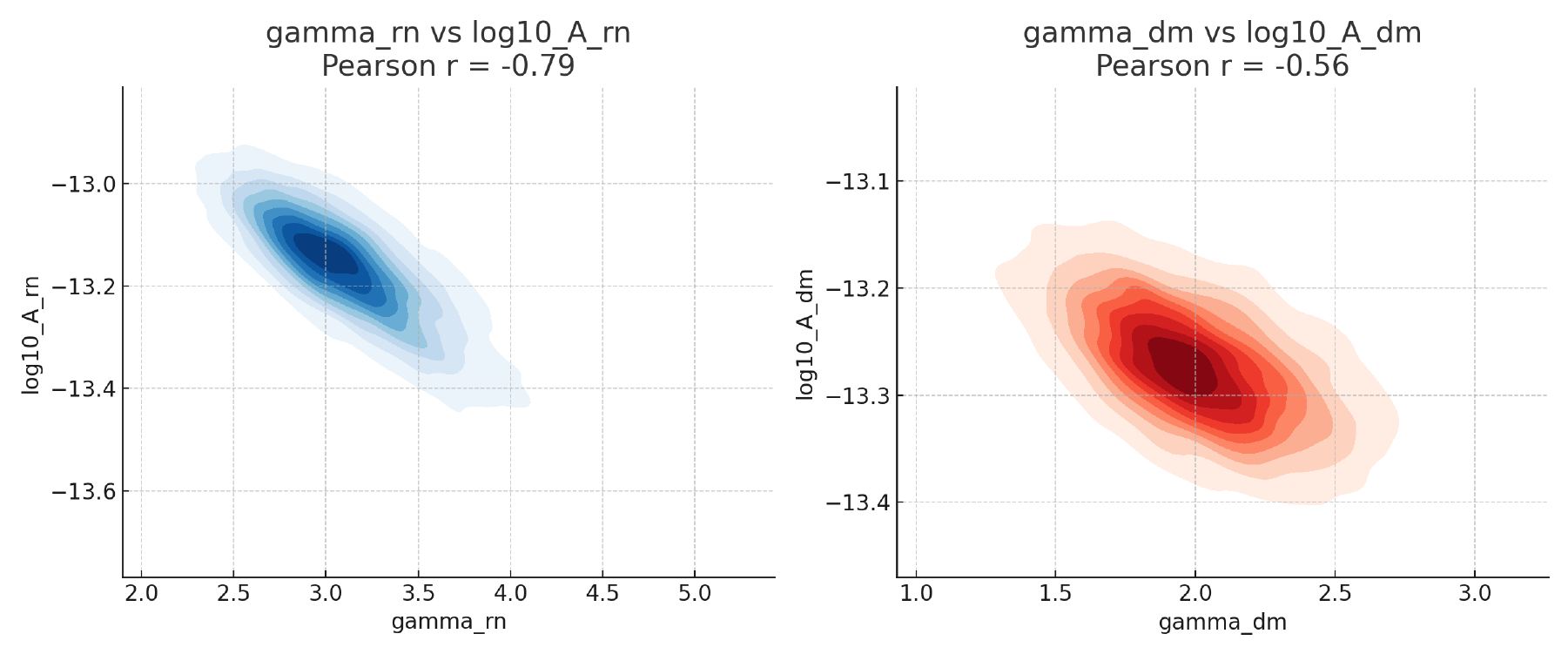} \label{fig:ellipses_norep}
  }
  \hfill
  \subfloat[
  Joint 2D posteriors for $\log_{10}A$ and $\gamma$ (RN, DM) after projection.
  ]{
    \includegraphics[width=0.45\textwidth]{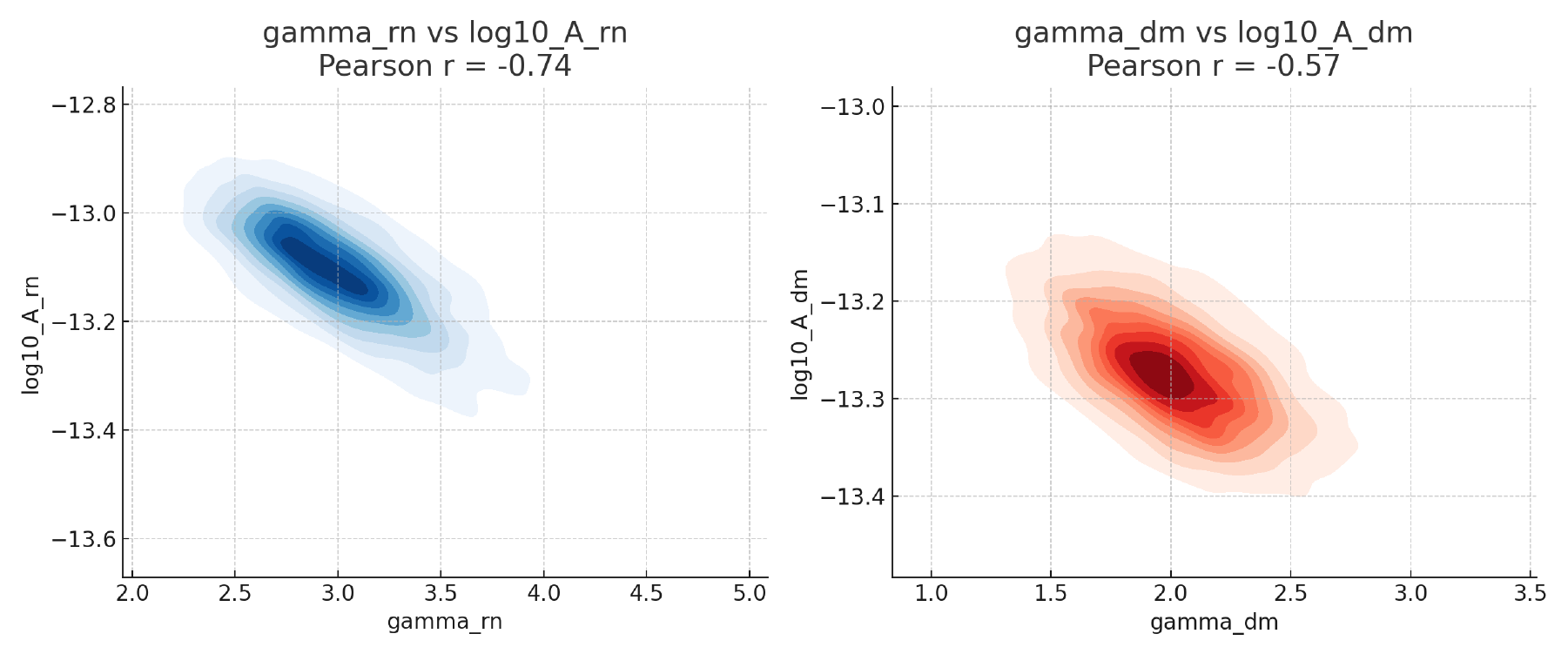}\label{fig:ellipses_rep_nobug}
  }
    \caption{
    Joint 2D posteriors for $\log_{10}A$ and $\gamma$ (RN, DM) with Gaussian prior-hyperprior before reparameterization; Pearson coefficient shown.
    }  \label{fig:ellipses}
\end{figure}

Figure~\ref{fig:correlation_original} shows the correlation structure in the original parameterization. While most parameter pairs exhibit high independence ($\mathcal{I} > 0.9$), notable exceptions include the coupling between $\log_{10}A_{\text{dm}}$ and the hyperparameters $\mu_{\gamma_{\text{rn}}}$ ($\mathcal{I} = 0.60$) and $\sigma_{\log_{10}A_{\text{rn}}}$ ($\mathcal{I} = 0.61$). These correlations reflect a well-known characteristic of power-law noise processes in PTA data analysis: the amplitude and spectral index parameters exhibit strong anticorrelation \cite{vanHaasteren2014}, \cite{Lentati2016}. This arises because, for a fixed dataset, a steeper spectrum (larger $\gamma$) can be partially compensated by a larger amplitude, creating a degeneracy in the likelihood surface. This anticorrelation is particularly pronounced for red noise, where typical values yield $\rho \approx -0.7$ to $-0.9$ between $\log_{10}A$ and $\gamma$ \cite{EPTA2023}. A similar but weaker anticorrelation exists for DM variation noise, as both processes share the same power-law spectral form. Crucially, these physical correlations persist even with hierarchical hyperprior structure, as evidenced by the moderate independence scores in Figure~\ref{fig:correlation_original} and the correlation in the 2D joint posterior distribution in Figure~\ref{fig:ellipses_norep}. See also Figures 1 and 2 in \cite{ensemble}. As expected, the hierarchical structure on the noise priors does not eliminate the inherent parameter degeneracies in the underlying modeling of the noise signal.

After applying the orthogonal projection to obtain $\tilde{\bm{\Lambda}}$ (Figure~\ref{fig:correlation_corrected}), we observe a mixed but revealing pattern of decorrelation. The projection successfully eliminates several specific correlations present in the original parameterization: the couplings between $\log_{10}A_{\text{dm}}$ and both $\mu_{\gamma_{\text{rn}}}$ and $\sigma_{\gamma_{\text{rn}}}$ improve from moderate correlation ($\mathcal{I} = 0.60$ and $0.77$, respectively) to complete independence ($\mathcal{I} = 1.00$). Similarly, the correlation between $\log_{10}A_{\text{dm}}$ and $\sigma_{\log_{10}A_{\text{rn}}}$ is fully removed (from $\mathcal{I} = 0.61$ to $1.00$). In contrast, the weak correlation between $\log_{10}A_{\text{dm}}$ and $\mu_{\log_{10}A_{\text{rn}}}$ remains largely unchanged (from $\mathcal{I} = 0.89$ to $0.83$), suggesting that this particular coupling is not addressed by the orthogonal projection. Most notably, the amplitude parameter $\log_{10}A_{\text{dm}}$ exhibits anomalous behavior with respect to its own transformed hyperparameters. This indicates that the orthogonal projection, rather than decorrelating these parameters, has concentrated approximately 81\% of the variance in $\log_{10}A_{\text{dm}}$ into its transformed mean hyperparameter. This selective failure — affecting primarily $\log_{10}A_{\text{dm}}$ while leaving other parameters successfully decorrelated — suggests that the issue is not systemic but rather specific to how the projection interacts with the amplitude parameter under Gaussian priors. The concentration of residual correlations in the DM amplitude parameters may reflect the combined effect of the inherent amplitude-spectral index anticorrelation in power-law processes and the constraints imposed by the Gaussian hyperprior structure. These results indicate that there is room for improving the reparametrization procedure, particularly in the way it handles pre-existing anticorrelations peculiar of the PTA data and hierarchical modeling choices. Let us first comment on our prior and hyperprior choices. In the initial implementation presented in this work, we chose to adopt Gaussian priors for both physical parameters and hyperpriors, with hyperparameters means and standard deviations taken from Table 2 in \cite{ensemble}, who performed inference on the hyperparameters directly. This choice makes our analysis methodologically robust, as it is grounded in estimates derived from the EPTA dataset. However, the effects after reparametrization may be partially attributable to our choice of Gaussian priors with relatively tight hyperpriors. As shown explicitly in Figure~\ref{fig:correlations}, the mean of $\log_{10}A_{\text{dm}}$ exhibits a particularly anomalous behavior: under Gaussian priors with means and variances from \cite{ensemble}, the hyperparameter $\mu_{\log_{10}A_{\text{dm}}}$ controls the mean of a relatively narrow distribution for $\log_{10}A_{\text{dm}}$. The orthogonal projection, in removing the component of $\mu_{\log_{10}A_{\text{dm}}}$ that lies in the subspace spanned by the physical parameters, may concentrate the remaining variation into a direction that is maximally aligned with $\log_{10}A_{\text{dm}}$ itself. This effect is enhanced by the tight Gaussian hyperpriors, which limit the available parameter space and make the transformed parameter $\tilde{\mu}_{\log_{10}A_{\text{dm}}}$ essentially a rescaled version of the physical parameter it was meant to be decorrelated from. This observation suggests that the combination of Gaussian priors and orthogonal projection may be particularly unsuitable for amplitude parameters in hierarchical PTA noise models. The failure could be specific rather than systemic: alternative prior specifications, particularly uniform priors on amplitudes or more flexible hyperprior distributions, may avoid this pathological behavior by providing more degrees of freedom that survive the projection. 

Let us finally comment on the fundamental correlations inherent to the PTA noise models. Figure~\ref{fig:ellipses_rep_nobug} shows that the characteristic anticorrelation between amplitude and spectral index parameters also persists after the projection. This preservation is crucial, as these anticorrelations are not artifacts of the hierarchical structure but rather reflect the intrinsic degeneracies in the power-law noise modeling. Despite the complex procedure involved in the reparametrization of the hyperparameters, our method correctly captures and maintains the structure of the modeling of the lower hierarchical level. This robustness is essential for ensuring that any gains from reparametrization do not come at the cost of losing underlying parameter relationships. The ability of the NFs-based approach to maintain these intrinsic correlations while attempting to decorrelate hierarchical dependencies represents both a strength and a challenge. On one hand, it demonstrates that the method does not artificially destroy the essential structure of the noise model, which would compromise the physical interpretability of the results. On the other hand, it highlights the fundamental difficulty in distinguishing between correlations that arise from the hierarchical prior structure (which we aim to mitigate) and those that are inherent to the modeling of the process (which must be preserved).

\section{Concluding remarks}\label{sec:concl}
In this paper, we have presented in full detail a hierarchical Bayesian framework for PTA noise analysis that systematically addresses prior dependence. The starting point is the introduction of hyperpriors on pulsar noise parameters: rather than using fixed priors on individual pulsar noise parameters, we introduced hyperpriors to describe the population-level distribution of these parameters, building up a hierarchical structure where hyperparameters govern the overall noise characteristics across the array of pulsars. We developed an orthogonal reparametrization strategy with the aim to address the correlations between physical parameters and hyperparameters. It is based on the employment of NFs, which provide flexible and tractable mappings between the two parameter spaces and model directly both the conditional distribution for the physical parameters and their hyperprior in the transformed parameter space.

The validation test in Section~\ref{sec:testA} confirms that combining hierarchical modeling in \texttt{Enterprise} with \texttt{i-nessai} sampling offers a consistent and statistically grounded framework to investigate how hyperpriors affect PTA inference. As a first application of our approach, we present the effects of the reparametrization on the RN and DM variation noise parameters for a single-pulsar with observed TOA simulated from the European Pulsar Timing Array dataset DR2new. 

In summary, our work shows that orthogonal projection provides a principled first step toward reducing prior dependence in hierarchical PTA models: while preserving intrinsic parameter correlations of the underlying noise modeling, it does not fully disentangle them from those arising from the hierarchical structure considered here. The residual dependencies observed indicate that further refinements are required, in particular: (i) the use of more flexible prior specifications, such as a uniform prior on the physical parameters, that could potentially perform better, since the Gaussian assumption may impose a rigid hierarchical structure that, when combined with the orthogonal projection, may overly constrain the transformed parameter space and (ii) improvements of the NFs-guided reparametrization that can explicitly differentiate between power-law modeling and hierarchical correlations, potentially through physics-informed neural network architectures or by incorporating domain knowledge directly into the flow design.

The use of NFs is ubiquitous in our work: they are employed not only to realize orthogonal reparametrization, but also within the sampling algorithm. Specifically, we adopted \texttt{i-nessai}, a flow-guided nested sampler that leverages NFs to accelerate exploration of high-dimensional and computationally expensive posterior distributions. This combination ensures that both the statistical formulation and the computational implementation are consistently supported by flow-based methods.

\begin{credits}
\subsubsection{\ackname} We would like to thank Marco Bonici for his valuable insights on the details of the approach used in \cite{Paradiso_2025} and for stimulating discussions on our work. This paper is supported by the  Fondazione  ICSC, Spoke 3 Astrophysics and Cosmos Observations. National Recovery and Resilience Plan (Piano Nazionale di Ripresa e Resilienza, PNRR) Project ID CN-00000013 "Italian Research Center on High-Performance Computing, Big Data and Quantum Computing" funded by MUR Missione 4 Componente 2 Investimento 1.4: Potenziamento strutture di ricerca e creazione di "campioni nazionali di R\(\&\)S (M4C2-19 )" - Next Generation EU (NGEU). All HPC tests and benchmarks of this work have been carried out on the cluster at the Osservatorio di Astrofisica e Scienza dello Spazio (OAS), Bologna (Italy). 
We are grateful to Alessandro Tacchini for providing us access to the computing cluster.

\end{credits}

%
%
%
\bibliographystyle{splncs04}
\bibliography{bibliography}

\end{document}